\documentstyle[aps,epsf]{revtex}
\begin{document}

\title {Quantum electromagnetic field in a three dimensional
oscillating cavity}
\author{Martin Crocce $^{1}$ \thanks{Hector.Crocce@physics.nyu.edu},
Diego A.\ R.\ Dalvit $^{2}$ \thanks{dalvit@lanl.gov},
and Francisco D.\ Mazzitelli $^{3}$ \thanks{fmazzi@df.uba.ar} }
\address{$^1$
Dept. of Physics, New York University, 4 Washington Pl., NY 10003}
\address{$^2$ Theoretical Division, MS B210, Los Alamos National
Laboratory, Los Alamos, NM 87545}
\address{$^3$
Departamento de F\'\i sica {\it J.J. Giambiagi},
Facultad de Ciencias Exactas y Naturales\\
Universidad de Buenos Aires - Ciudad Universitaria, Pabell\' on I,
1428 Buenos Aires, Argentina}
\maketitle

\begin{abstract}

We compute the photon creation inside a perfectly conducting,
 three dimensional oscillating cavity, taking the polarization of the
electromagnetic field into account.
As the boundary conditions for this field are both of
Dirichlet and (generalized) Neumann type, we analyze as a preliminary
step the dynamical Casimir effect for a scalar field satisfying
generalized  Neumann boundary conditions. We show
that particle production is enhanced with respect to the case of
Dirichlet boundary conditions. Then we consider
the transverse electric and transverse magnetic polarizations of
the electromagnetic field.
For resonant frequencies, the total number of photons grows exponentially
in time for both polarizations, the rate being greater for
transverse magnetic modes.

\end{abstract}


%%%%%%%%%%%%%%%%%%%%%%%%%%%%%%%%%%%%%%%%%%%%%%%%%%%%%%%%%%%%%%%%%%%%%%%%%%%%%%%

\section{Introduction}

The existence of an attractive force between two uncharged,
perfectly conducting parallel plates was predicted
by Casimir in 1948 \cite{Casimir} and has recently been measured at
the 15\% precision level using state-of-the-art cantilevers \cite{roberto}.
A similar force between a conducting plane
and a sphere  has also been measured with progresively higher precision in the
last years using torsion balances \cite{Lamoreaux}, atomic force microscopes
\cite{Mohideen}, and capacitance bridges \cite{chan1,chan2}, with
the latter reference showing the relevance of Casimir forces
in nanotechnology. For a recent review of
experimental and theoretical developments see \cite{Bordag}.

The dynamical effect consists in the generation of photons due to the
instability of the vacuum state of the electromagnetic field
in the presence of time-dependent boundaries. In the
literature it is referred to as dynamical Casimir effect \cite{Schwinger}
or motion-induced radiation \cite{Lambrecht}. The dynamical effect
has been recently reviewed in Ref. \cite{Dodonov-rev}.
Up to now no concrete experiment has been performed
to confirm this photon generation, but an experimental
verification is not out of reach. From 
the theoretical point of view it is widely accepted that the most
favorable configuration in order to observe the phenomenon is a
vibrating cavity in which it is possible to  produce resonant effects
between the mechanical and field oscillations.

Many previous papers have focused their attention in the
scalar field quantization within a one dimensional cavity \cite{1-dim,dodexp}.
Recently, we analyzed in detail the case of a three dimensional
cavity \cite{crocce},
but still considering a scalar field (in this
and other \cite{more} previous works
Dirichlet boundary
conditions are assumed). The main difference between one and three dimensional
cavities is that, while in one dimension
the cavity's frequency spectrum is equidistant and leads to
strong intermode interactions, in three
dimensions the spectrum is in general nonequidistant, and only
a few modes may be coupled. The relevance of this  coupling
has been pointed out only recently (see \cite{crocce,Dodonov2}) .
The aim of this paper is to extend the results
of Ref. \cite{crocce} to the realistic case of the electromagnetic field,
properly taking into account the polarization of the different
modes (transverse electric (TE) and transverse magnetic (TM) polarizations).

As we will see in Sec. II, the electromagnetic field involves both
Dirichlet and
(generalized) Neumann boundary conditions. For this reason, it is
of interest to
analyze the case of a massless scalar field satisfying this
latter type of boundary
conditions, which we do in Sec. III. Assuming a resonant vibration
of the cavity and using multiple scale analysis (MSA) \cite{Bender}
we will show that the number of particles produced is much larger
than for Dirichlet
boundary conditions. We study in detail the resonant case in which the cavity
oscillates at twice the frequency of some field mode.
In Sec. IV we show
that TE modes of the electromagnetic field
behave as a scalar field with Dirichlet boundary
conditions, while TM modes are analogous to the scalar
case with Neumann boundary conditions. Section V contains our main conclusions.

%%%%%%%%%%%%%%%%%%%%%%%%%%%%%%%%%%%%%%%%%%%%%%%%%%%%%%%%%%%%%%%%%%%%%%%%%%%%%%

\section{The boundary conditions}

We consider a rectangular cavity formed by perfectly reflecting walls
with dimensions $L_{x},L_{y},$ and $L_{z}$. The wall placed at $x=L_{x}$
is at rest for $t<0$ and begins to move following a given trajectory,
$L_{x}(t)$, at $t=0$. We assume this trajectory as prescribed for
the problem (not a dynamical variable) and that it works as a time-dependent
boundary condition for the field. Moreover, we will assume a non
relativistic motion of the wall with $L_{x}(t)=L_0(1+\epsilon f(t))$
with $\epsilon\ll 1$ and $f(t)$ a bounded function. We use units
$\hbar=c=1$.

Let us consider the problem of finding the electromagnetic field
inside the cavity in
terms of the four vector potential $A_{\mu}=(\varphi,{\bf A})$.
In the Coulomb gauge $\nabla \cdot{\bf A} = 0$
the scalar potential $\varphi$ vanishes
and the vector potential satisfies the wave equation
$\Box{\mathbf{A}}=0$. For the
static walls, the boundary conditions are the usual ones

\begin{equation}
\mathbf{E}_{\parallel}=0 \ \ \ ; \ \ \ \mathbf{B}_{\perp}=0,
\label{cond. de contorno}
\end{equation}
where  $\parallel$ and $\perp$ respectively denote the components of the field
parallel and perpendicular to the wall.

On the moving
wall, these boundary conditions must be imposed in a Lorentz frame
in which the mirror is instantaneously at rest. As the mirror moves
in the $x$ direction, it will be convenient to decompose
the electromagnetic fields into  TE and TM modes with respect to
the $x$ axis.
The TE fields are defined as the solutions to Maxwell equations
with
${\mathbf E}^{( {\rm TE})} \cdot {\mathbf \hat x}=0$. Analogously,
the TM fields satisfy ${\mathbf B}^{( {\rm TM})} \cdot {\mathbf \hat x}=0$.
A general solution for the electromagnetic field inside the cavity
is a superposition of TE and TM fields.

It is useful to introduce a {\it different} vector potential for
each polarization through the equations \cite{Netojpa,Neto}

\begin{eqnarray}
{\mathbf E}^{({\rm TE})} &=& -\partial_t {\mathbf A}^{({\rm TE})}
\,\, ; \,\, {\mathbf B}^{({\rm TE})} =
\nabla\times {\mathbf A}^{({\rm TE})} ; \\
{\mathbf B}^{({\rm TM})} &=& \partial_t{\mathbf A}^{({\rm TM})}
\,\,\, ; \,\,\, {\mathbf E}^{({\rm TM})} =
\nabla\times{\mathbf A}^{({\rm TM})} .
\label{TETM}
\end{eqnarray}
Both potentials satisfy the Coulomb gauge and have vanishing $x$ component.
As ${\mathbf A} \cdot {\mathbf \hat x} = 0$ and $\varphi = 0$, the
vector potentials are invariant
under a boost in the $x$-direction. The same is true for the Coulomb gauge.
In terms of these potentials, the boundary conditions are relatively simple
\cite{Neto}.
Let us denote by $S$ the laboratory frame and by $S'$ the instantaneous
comoving frame. In $S'$ the TE vector potential satisfies Dirichlet
boundary conditions
${\mathbf A}'^{({\rm TE})}(x'=0,y',z',t')={\mathbf 0}$. Therefore, on
the moving mirror,

\begin{equation}
{\mathbf A}^{({\rm TE})}(x=L_x(t),y,z,t)={\mathbf 0} .
\label{ATE}
\end{equation}
On the other hand, the TM vector potential satisfies

\begin{equation}
n^{\mu '}\partial_{\mu '}{\mathbf A}'^{({\rm TM})}(x'=0,y',z',t')=
{\mathbf 0} ,
\end{equation}
where $n^{\mu '}=(0,1,0,0)$. As a consequence, for a non relativistic
motion of the mirror,

\begin{equation}
n^{\mu }\partial_{\mu} {\mathbf A}^{({\rm TM})}(x=L_x(t),y,z,t) =
(\partial_x+\dot L_x(t)\partial_t)
{\mathbf A}^{({\rm TM})}(x=L_x(t),y,z,t) = {\mathbf 0} ,
\label{ATE2}
\end{equation}
i.e., a ``generalized" Neumann boundary condition with $n^{\mu}=
(\dot L_x,1,0,0)$.
On the static mirrors the boundary conditions for the TE vector potential
are given by

\begin{eqnarray}
&&  {\bf A}^{({\rm TE})}(x=0,y,z,t)= {\bf 0}, \nonumber \\
&&  A_y^{({\rm TE})}(x,y,\{z=0,L_z\},t)=
A_z^{({\rm TE})}(x,\{y=0,L_y\},z,t)= 0.
\label{condicionesTE}
\end{eqnarray}
For the TM potential we have

\begin{eqnarray}
&& \partial_x A_z^{({\rm TM})}(x=0,y,z,t)=
\partial_x A_y^{({\rm TM})}(x=0,y,z,t) = 0 , \nonumber \\
&& A_y^{({\rm TM})}(x,\{y=0,L_y\},z,t)=
\partial_y A_z^{({\rm TM})}(x,\{y=0,L_y\},z,t) = 0, \nonumber \\
&& A_z^{({\rm TM})}(x,y,\{z=0,L_z\},t)=
\partial_z A_y^{({\rm TM})}(x,y,\{z=0,L_z\},t) = 0.
\label{condicionesTM}
\end{eqnarray}
From these boundary conditions it is clear that the behavior of
each component
of the TE vector field is related to the problem of a scalar field subjected
to Dirichlet boundary conditions. For the TM vector field it is necessary
to deal with the generalized Neumann boundary conditions given in Eq.(\ref{ATE2}).
The former problem was
extensively studied in our previous paper \cite{crocce}, and the
latter one will be treated in the following section.

%%%%%%%%%%%%%%%%%%%%%%%%%%%%%%%%%%%%%%%%%%%%%%%%%%%%%%%%%%%%%%%%%%%%%%%%%%%%

\section{Scalar field with Neumann boundary conditions}

Let us consider the problem of a massless scalar field
$\phi({\mathbf x},t)$ satisfying the wave equation
$\Box\phi=0$ and (generalized) Neumann boundary conditions
on all surfaces of the cavity. In the comoving frame
the Neumann boundary condition is  $n^{\mu '}\partial_{\mu '}\phi '= 0$.
In the laboratory frame, this condition becomes $n^{\mu}\partial_{\mu}\phi
= 0$, where
$n^{\mu}=(\dot L_x,1,0,0)$. Therefore we have

\begin{eqnarray}
\partial_x \phi (x=0,y,z,t)=0 & \; ; \; & (\partial_x + \dot{L}_x \partial_t)
\phi(x=L_x(t),y,z,t) =0 ; \nonumber \\
\partial_y \phi(x,\{y=0,L_y\},z,t)=0 & \; ; \; &
\partial_z \phi(x,y,\{z=0,L_z\},t)=0.
\label{ncphi}
\end{eqnarray}

\subsection{Instantaneous basis}

The Fourier expansion of the field for an arbitrary moment of time
can be written in terms of creation and annihilation operators as

\begin{equation}
\phi({\bf x},t)=\sum_{{\bf n}} a_{{\bf n}}^{\scriptscriptstyle{\rm in}}
u_{{\bf n}}({\bf x},t) + {\rm H.c.} ,
\label{field}
\end{equation}
where the mode functions $u_{{\bf n}}({\bf x},t)$ form a complete
orthonormal set of solutions
of the wave equation with  Neumann boundary conditions.
When $t\leq0$ (static cavity) each field mode is determined by three
non-negative integers $n_{x},n_{y}$ and $n_{z}$. Namely

\begin{equation}
u_{{\bf{n}}}({\bf{x}},t<0)={1\over\sqrt{2\omega_
{\bf{n}}}}\sqrt{\frac{2}{L_{x}}}\cos\left(\frac{n_{x}\pi}
{L_{x}} x\right)\sqrt{\frac{2}{L_{y}}}
\cos\left(\frac{n_{y}\pi}{L_{y}} y\right)
\sqrt{\frac{2}{L_{z}}}\cos\left(\frac{n_{z}\pi}{L_{z}} z\right)
e^{- i\omega_{\bf{n}}t} ;
\label{expest}
\end{equation}
\begin{equation}
\omega_{\bf{n}}=\pi\sqrt{\left(\frac{n_{x}}{L_{x}}\right)^{2} \!+
\left(\frac{n_{y}}{L_{y}}\right)^{2}\! +
\left(\frac{n_{z}}{L_{z}}\right)^{2}} ,
\label{omega}
\end{equation}
with the shorthand ${\bf n}=(n_{x},n_{y},n_{z})$.
In order to satisfy the boundary conditions for $t>0$
it is useful to expand the mode
functions in Eq.(\ref{field}) with respect to an \textit{instantaneous basis}.
If the boundary condition on the moving mirror were the instantaneous
Neumann condition  $ \partial_x
\phi(x=L_x(t),y,z,t) =0$, the trivial choice for  the instantaneous
basis would be
\begin{equation}
\sqrt{\frac{2}{L_{x}(t)}}
\cos\left(\frac{k_{x}\pi}{L_{x}(t)} x\right)
\sqrt{\frac{2}{L_{y}}}\cos\left(\frac{k_{y}\pi}{L_{y}} y\right)
\sqrt{\frac{2}{L_{z}}}\cos\left(\frac{k_{z}\pi}{L_{z}} z\right) .
\end{equation}
However, as the generalized Neumann condition in Eq.(\ref{ncphi})
involves the time derivative of the field, the situation is more complex.

Let us consider new variables $(\eta,\xi)$ in the $(t,x)$ plane. We
define the line $\eta = {\rm const} $ as a slight modification of the
corresponding $ t = {\rm const}$ line, in such a way that it is orthogonal
to the worldlines of the mirrors at $x=0$ and $x=L_x(t)$ (see Figure).
The variable $\xi$ is defined as the distance, on the line $\eta={\rm const}$,
from $x=0$ to $x$. In these coordinates,
the generalized boundary condition on the two mirrors becomes
$ \partial_{\xi} \phi(\xi,y,z,\eta) =0$ both at $\xi = 0$ and at
$\xi = l(\eta)$, where $l(\eta)$ is the value of the coordinate $\xi$ on the
moving mirror.
Therefore, an
instantaneous basis to describe the field is

\begin{equation}
\sqrt{\frac{2}{l (\eta)}}
\cos\left(\frac{k_{x}\pi}{l(\eta)} \xi\right)
\sqrt{\frac{2}{L_{y}}}\cos\left(\frac{k_{y}\pi}{L_{y}} y\right)
\sqrt{\frac{2}{L_{z}}}\cos\left(\frac{k_{z}\pi}{L_{z}} z\right) .
\label{biv}
\end{equation}

\begin{figure}
\centering \leavevmode
\epsfxsize=5cm
\epsfbox{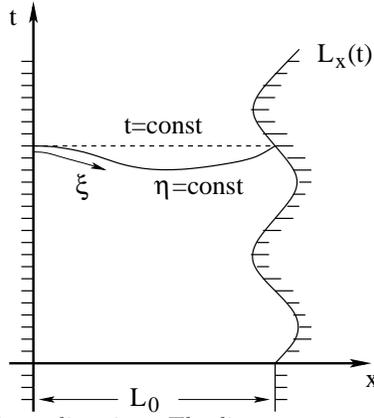}
\caption{Worldlines of the mirrors along the x direction. The line
$\eta={\rm const}$ is orthogonal to the worldlines at $x=0$ and $x=L_x(t)$,
and the coordinate $\xi$ measures the distance from the static mirror along
the $\eta={\rm const}$ line.}
\end{figure}

To find a concrete form for the new coordinates we write
$\eta=t + g(x,t)$. Therefore $\xi$ is given by

\begin{equation}
\xi=\int_0^x dx'\sqrt{1+{g'^2(x',t)\over (1+\dot g(x',t))^2}} .
\end{equation}

At this point it is important to note that, since we are considering
motions of the wall that are small ($O(\epsilon)$) deviations from the
initial static position, terms of order $O(\epsilon^2)$ or higher will be
neglected in what follows.
Moreover, $g(x,t)=O(\epsilon)$, 
$\xi = x + O(\epsilon^2)$, and $l(\eta)=L_x(t)+ O(\epsilon^2)$.
With this in mind, it is easy to show that,
in order to fulfill the assumed orthogonality between the line
$\eta = {\rm const}$ and the mirrors' worldlines, the function $g(x,t)$
must satisfy
\begin{eqnarray}
g(x=0,t)=0 & \; ; \; & g(x=L_x(t),t)=0 ; \nonumber \\
\partial_x g(x=0,t) = 0 & \; ; \; & \partial_x g(x=L_x(t),t) = - \dot{L} .
\label{bc}
\end{eqnarray}
There are many solutions to the above conditions,
that can be written in the form
\begin{equation}
g(x,t)= \dot{L}_x(t)  L_x(t) v(x/L_x(t)) ,
\end{equation}
where $v(0)=v(1)=0$, $v'(0)=0$ and $v'(1)=-1$ (the prime denotes
derivation with
respect to $x$). For example, a possible
solution is
$v (z) = \frac{1}{2} (z^2 - z^4)$.
Physical quantities like the number of created particles or
the energy density inside the cavity should not depend on the choice
of $g(x,t)$. We will keep  a general function as a benchmark for our
calculations.

Finally, the mode functions in Eq.(\ref{field}) can be expanded
in terms of the instantaneous basis Eq.(\ref{biv}) as follows
\begin{eqnarray}
u_{{\bf n}}({\bf x},t>0)&=&
\sum_{{\mathbf k}}
 Q_{\mathbf{k}}^{(\mathbf{n})}(\eta)
\sqrt{\frac{2}{l(\eta)}}
\cos\left(\frac{k_{x}\pi}{l(\eta)} \xi\right)
\sqrt{\frac{2}{L_{y}}}\cos\left(\frac{k_{y}\pi}{L_{y}} y\right)
\sqrt{\frac{2}{L_{z}}}\cos\left(\frac{k_{z}\pi}{L_{z}} z\right)
\nonumber\\
&\simeq&
\sum_{{\mathbf k}}
[ Q_{\mathbf{k}}^{(\mathbf{n})}(t) +
\dot{Q}_{\mathbf{k}}^{(\mathbf{n})}(t) g(x,t) ]
\sqrt{\frac{2}{L_{x}(t)}}
\cos\left(\frac{k_{x}\pi}{L_{x}(t)} x\right)
\sqrt{\frac{2}{L_{y}}}\cos\left(\frac{k_{y}\pi}{L_{y}} y\right)
\sqrt{\frac{2}{L_{z}}}\cos\left(\frac{k_{z}\pi}{L_{z}} z\right)\nonumber\\
& \equiv &\sum_{{\bf k}}
[ Q_{{\bf k}}^{({\bf n})}(t) + \dot{Q}_{{\bf k}}^{({\bf n})}(t) g(x,t) ]
\, \phi_{{\bf k}}({\bf x},L_{x}(t)) .
\label{exp}
\end{eqnarray}
The initial conditions are given by
\begin{equation}
Q_{\bf{k}}^{(\bf{n})}(0)={1\over\sqrt{2\omega_{\bf{n}}}}\,
\delta_{\bf{k},\bf{n}}  \ , \ \
\dot{Q}_{\bf{k}}^{(\bf{n})}(0)=-i
\sqrt{\omega_{\bf{n}}\over2}\,\delta_{\bf{k},\bf{n}} ,
\end{equation}
provided that $L_{x}(t)$ and $\dot{L}_{x}(t)$ are continuous at $t=0$,
and that the initial acceleration $\ddot{L}_x (0)= O(\epsilon^2)$.
In this way we ensure that each field mode and its time derivative are
also continuous functions at $t=0$.

\subsection{Dynamical equations}

We now study the trajectory $L_x(t) = L_0 (1+\epsilon \sin(\Omega t))$.
{\footnote{Strictly speaking, we should add to $L_x(t)$
some decaying function in order to
meet the continuity conditions at $t=0$. Since we will be interested in a
resonant behavior of the field, this additional function will not contribute,
being irrelevant for what follows. For a more detailed discussion of this
point see \cite{crocce}.}}
The equations for the modes
$Q_{\bf{k}}^{\bf{n}}(t)$ can be obtained from Eq.(\ref{exp}), since
$\Box u_{{\bf n}}({\bf x},t>0) = 0$. We first apply the Dalambertian, and then
multiply by $\phi_{{\bf k}}$ and integrate over the
cavity. To order
$O(\epsilon^2)$, the equations read

\begin{eqnarray}
\ddot{Q}_{\bf{k}}^{(\bf{n})}+\omega_{\bf{n}}^{2}(t)\, Q_{\bf{k}}^{(\bf{n})}
&=&
- 2\lambda(t)\sum_{\bf{j}} g_{\bf{jk}}  \dot{Q}_{\bf{j}}^{(\bf{n})}
- \dot{\lambda}(t) \sum_{\bf{j}}g_{\bf{jk}} Q_{\bf{j}}^{(\bf{n})} \nonumber \\
&& - 2 \dot{\lambda}(t) L_x^2(t) \sum_{\bf{j}} r_{\bf{jk}}
\ddot{Q}_{\bf{j}}^{(
\bf{n})}
- \sum_{\bf{j}} \dot{Q}_{\bf{j}}^{(\bf{n})} \;
(r_{\bf{jk}} \ddot\lambda(t)L_x^2(t)  - \lambda(t) \eta_{\bf{jk}})
- \lambda(t)L_x^2(t)\sum_{\bf{j}} r_{\bf{jk}}  \partial_t
\ddot{Q}_{\bf{j}}^{(\bf{n})} ,
\label{ecq}
\end{eqnarray}
where

\begin{eqnarray}
\omega_{\mathbf{k}}(t)&=&\pi\sqrt{\left(\frac{k_{x}}{L_{x}(t)}\right)^{2} \,+
\left(\frac{k_{y}}{L_{y}}\right)^{2}\, +
\left(\frac{k_{z}}{L_{z}}\right)^{2}} ,
\\
\lambda(t) &=& \frac{\dot{L}_{x}(t)}{L_{x}(t)} , \\
r_{\bf{jk}} &=& \int_0^{L_x(t)} dx \int_0^{L_y} dy \int_0^{L_z} dz \; v
\phi_{\bf j} \phi_{\bf k} , \\
\eta_{\bf{jk}} &=& L^2_x(t) \int_0^{L_x(t)} dx \int_0^{L_y} dy
\int_0^{L_z} dz \;
[ (v^{''}- \omega^2_{\bf j} v) \phi_{\bf j} \phi_{\bf k}
+ 2 v^{'} \phi_{\bf j}^{'} \phi_{\bf k} ] .
\end{eqnarray}
Here, $\omega_{\bf j}$ is the frequency of the mode evaluated at
$\epsilon=0$. As before,
the prime denotes derivation with respect to $x$. The coefficients
$g_{\bf{jk}}$ are defined by

\begin{equation}
g_{\bf{jk}}= L_{x}(t)\int_{0}^{L_{x}(t)}dx \int_0^{L_y} dy \int_0^{L_z} dz
\frac{\partial \phi_{\bf{j}}}{\partial L_{x}}\,\phi_{\bf{k}} =
\left\{
\begin{array}{ll}
(-1)^{k_{x}+j_{x}} \frac{2 j_{x}^2 }{k_{x}^{2}-j_{x}^{2}}\,
\delta_{k_{y}j_{y}}\,\delta_{k_{z}j_{z}} \,& \mbox{if $k_{x}\neq j_{x}$} ,\\
- \delta_{k_{y}j_{y}}\,\delta_{k_{z}j_{z}}
 & \mbox{if $k_{x}=j_{x}$}  .
\end{array}
\right.
\label{ge}
\end{equation}

Had we considered Dirichlet boundary conditions on the
static walls ($y=0,L_y;z=0,L_z$) we would had obtained the same dynamical
equations for the modes $Q_{\bf{k}}^{\bf{n}}(t)$; i.e., the form of
the equations only depends on the boundary conditions imposed along the
$x$-direction. This is because the coefficients
$r_{\bf{jk}}, \eta_{\bf{jk}}$ and $g_{\bf{jk}}$ do not depend on the
particular form of the $\phi_{\bf{k}}$ in the plane $y-z$, as long as
they are properly normalized in this plane.
However, when Dirichlet boundary conditions on
$x=0$ and $x=L_x(t)$ are considered, the equations for the modes
are different (see Ref. \cite{crocce} and Eqs.(\ref{ecqte}) and (\ref{gdir})
below. Note that
the coefficients $g_{{\bf j k}}$ for Neumann boundary
conditions are different from those for Dirichlet boundary
conditions.

When the mirror returns to its initial position  for $t>t_{\rm final}$
the rhs of Eq.(\ref{ecq}) vanishes and the solution reads

\begin{equation}
Q_{\mathbf{k}}^{(\mathbf{n})}(t>t_{\rm final})=
A_{\mathbf{k}}^{(\mathbf{n})}e^{i\omega_{\mathbf{k}}t}+
B_{\mathbf{k}}^{(\mathbf{n})}e^{-i\omega_{\mathbf{k}}t},
\label{sol}
\end{equation}
with $A_{\mathbf{k}}^{(\mathbf{n})}$ and $B_{\mathbf{k}}^{(\mathbf{n})}$
being some constant coefficients to be determined by the continuity
conditions at $t=t_{\rm final}$. The number of particles in the mode
${\bf k}$ is given by

\begin{equation}
\langle {\mathcal{N}}_{{\bf k}} \rangle=\sum_{{\bf n}}2\omega_
{\mathbf{k}}|A_{{\bf k}}^{({\bf n})}|^{2}
\label{numerodefotoneselectromagneticos}.
\end{equation}

%%%%%%%%%%%%%%%%%%%%%%%%%%%%%%%%%%%%%%%%%%%%%%%%%%%%%%%%%%%%%%%%%%%%

\subsection{Multiple scale analysis}

In order to find a solution to Eq.(\ref{ecq}) we use the multiple scale
analysis technique \cite{Bender}, which we have already applied in our
previous paper \cite{crocce}. We first introduce a second timescale
$\tau = \epsilon t$ and expand
$Q_{{\bf k}}^{({\bf n})} = Q_{{\bf k}}^{({\bf n})(0)} +
\epsilon  Q_{{\bf k}}^{({\bf n})(1)}$.
Replacing this into Eq.(\ref{ecq}) we obtain, as zeroth order solution,
 $Q_{{\bf k}}^{({\bf n})(0)} = A_{{\bf k}}^{({\bf n})}(\tau)
e^{i \omega_{{\bf k}} t} +
B_{{\bf k}}^{({\bf n})}(\tau) e^{-i \omega_{{\bf k}} t}$.
The functions $A_{{\bf k}}^{({\bf n})}(\tau)$ and
$B_{{\bf k}}^{({\bf n})}(\tau)$ will be obtained from imposing that no
secularities appear in the equation for
$Q_{{\bf k}}^{({\bf n})(1)}$. This reads

\begin{eqnarray}
\ddot Q^{({\bf n})(1)}_{\bf k} + \omega^2_{\bf k}
Q^{({\bf n})(1)}_{\bf k}
&=&
-2 \partial_{\tau } \dot Q^{({\bf n})(0)}_{\bf k} +
2 \left(\frac{\pi k_x}{L_0}\right)^2 \sin(\Omega t) Q^{({\bf n})(0)}_{\bf k}
\nonumber \\
&& + 2 \Omega \cos(\Omega t) \dot{Q}^{({\bf n})(0)}_{\bf k}
- \Omega^2 \sin(\Omega t) Q^{({\bf n})(0)}_{\bf k} \nonumber \\
&& - 2 L_0^2 \Omega^2 \sin(\Omega t) r_{\bf{kk}}\omega^2_{{\bf k}}
Q^{({\bf n})(0)}_{\bf k}
+ L_0^2 \Omega \cos(\Omega t) r_{\bf{kk}}\omega^2_{{\bf k}}
\dot{Q}^{({\bf n})(0)}_{\bf k} \nonumber \\
&& + L_0^2 \Omega \cos(\Omega t) \dot{Q}^{({\bf n})(0)}_{\bf k}
[ \Omega^2  r_{\bf{kk}} + \frac{1}{L_0^2} \eta_{\bf{kk}} ] \nonumber \\
&& + \Omega^2 \sin(\Omega t) \sum_{{\bf j} \neq {\bf k}} g_{\bf jk}
Q^{({\bf n})(0)}_{\bf j}
- 2 \Omega \cos(\Omega t) \sum_{{\bf j} \neq {\bf k}} g_{\bf jk}
\dot{Q}^{({\bf n})(0)}_{\bf j} \nonumber \\
&& - 2 L_0^2 \Omega^2 \sin(\Omega t) \sum_{{\bf j} \neq {\bf k}}
r_{\bf jk} \omega^2_{{\bf j}}  Q^{({\bf n})(0)}_{\bf j}
+ L_0^2 \Omega \cos(\Omega t)  \sum_{{\bf j} \neq {\bf k}}
r_{\bf jk}\omega^2_{{\bf j}}  \dot{Q}^{({\bf n})(0)}_{\bf j} \nonumber \\
&& + L_0^2 \Omega \cos(\Omega t) \sum_{{\bf j} \neq {\bf k}}
\dot{Q}^{({\bf n})(0)}_{\bf j}
\left[ \Omega^2 r_{\bf jk} + \frac{1}{L_0^2} \eta_{\bf{jk}} \right] .
\label{msa1}
\end{eqnarray}
where we have used that, to zeroth order, $\ddot {Q}^{({\bf n})(0)}_{\bf k}
= - \omega^2_{{\bf k}} Q^{({\bf n})(0)}_{\bf k}$.

The equations for $A_{{\bf k}}^{({\bf n})}(\tau)$ and
$B_{{\bf k}}^{({\bf n})}(\tau)$ are obtained imposing the condition
that any term in the right-hand side of Eq.(\ref{msa1}) with a time dependency
of the form $e^{\pm i \omega_{{\bf k}} t}$ must vanish. We get

\begin{eqnarray}
\frac{d A^{({\bf n})}_{\bf k}}{d\tau} &=&
-\frac{1}{2 \omega_{\bf k}} \left[
\frac{k_x^2 \pi^2}{L_0^2} - 2 \omega_{\bf k}^2 \right]
B^{({\bf n})}_{\bf k} \delta(2 \omega_{\bf k} - \Omega) \nonumber \\
&&
+ \sum_{\bf j \neq k} \left[
-(-\omega_{\bf j} + \frac{\Omega}{2}) g_{\bf jk} +
\delta_{k_y j_y} \delta_{k_z j_z} \omega_{\bf j} \right]
\delta(-\omega_{\bf k} - \omega_{\bf j} + \Omega)
\frac{\Omega}{2 \omega_{\bf k}}  B^{({\bf n})}_{\bf j} \nonumber \\
&& + \sum_{\bf j \neq k} \left[
-(\omega_{\bf j} + \frac{\Omega}{2}) g_{\bf jk} -
\delta_{k_y j_y} \delta_{k_z j_z} \omega_{\bf j} \right]
\delta(\omega_{\bf k} - \omega_{\bf j}  - \Omega)
\frac{\Omega}{2 \omega_{\bf k}}  A^{({\bf n})}_{\bf j} \nonumber \\
&& + \sum_{\bf j \neq k} \left[-
(\omega_{\bf j} - \frac{\Omega}{2}) g_{\bf jk} -
\delta_{k_y j_y} \delta_{k_z j_z} \omega_{\bf j} \right]
\delta(\omega_{\bf k} - \omega_{\bf j}  + \Omega)
\frac{\Omega}{2 \omega_{\bf k}}  A^{({\bf n})}_{\bf j} ,
\label{derA}
\end{eqnarray}
and an analogous equation for $B^{({\bf n})}_{\bf k}$, obtained by
the interchange $A^{({\bf n})}_{\bf k} \leftrightarrow
B^{({\bf n})}_{\bf k}$. Note that Eq.(\ref{derA}) is independent of
$g(x,t)$. This non trivial check of our calculations  follows
from two identities we used to derive Eq.(\ref{derA}),
namely

\begin{eqnarray}
-\frac{1}{2 \omega_{\bf k}} L_0^2 \omega^2_{\bf k} B^{({\bf n})}_{\bf k}
\delta(2 \omega_{\bf k} - \Omega)
\int_0^{L_x(t)} dx \int_0^{L_y} dy \int_0^{L_z} dz
(\phi^2_{\bf k} v^{'})^{'} &=&  \omega_{\bf k}  B^{({\bf n})}_{\bf k}
\delta(2 \omega_{\bf k} - \Omega) , \\
\omega_{\bf k}^2 r_{\bf jk} + \frac{1}{L_0^2} \eta_{\bf jk} =
\int_0^{L_x(t)} dx \int_0^{L_y} dy \int_0^{L_z} dz
[v' \phi_{\bf j} \phi_{\bf k} + v (\phi_{\bf k} \phi_{\bf j}^{'} -
\phi_{\bf k}^{'} \phi_{\bf j} )]'
&=& - \frac{2}{L_0^2} \delta_{k_y j_y} \delta_{k_z j_z} ,
\end{eqnarray}
where we have used the boundary conditions $v'(0)=0$ and $v'(1)=-1$.

%%%%%%%%%%%%%%%%%%%%%%%%%%%%%%%%%%%%%%%%%%%%%%%%%%%%%%%%%%%%%%%%%%

\subsection{Examples}

Let us consider the ``parametric resonant case'',
in which the external frequency $\Omega$ is twice the
frequency of an unperturbed mode ${\bf {k}}$, $\Omega=2\omega_{\bf k}$.
A second mode
${\bf {j}}$ will be coupled to the mode
${\bf {k}}$ iff $\vert \omega_{\bf k}\pm
\omega_{\bf j}\vert =\Omega $. We first assume this is not
the case. Therefore the evolution
equations become

\begin{eqnarray}
{d A^{({\bf n})}_{\bf k}\over d\tau} &=&
-\frac{1}{2 \omega_{\bf k}} \left[
\frac{k_x^2 \pi^2}{L_0^2} - 2 \omega_{\bf k}^2 \right]
B^{({\bf n})}_{\bf k} , \nonumber\\
{d B^{({\bf n})}_{\bf k}\over d\tau} &=&
-\frac{1}{2 \omega_{\bf k}} \left[
\frac{k_x^2 \pi^2}{L_0^2} - 2 \omega_{\bf k}^2 \right]
A^{({\bf n})}_{\bf k} .
\label{uncoupled}
\end{eqnarray}
It is easy to check from these equations  that $A^{({\bf n})}_{\bf k}$ and
$B^{({\bf n})}_{\bf k}$ grow exponentially as $e^{\lambda_N\tau}$, with a rate
$\lambda_N={1\over 2\omega_{\bf k}}(\omega_{\bf k}^2+\omega_{p}^2)$,
where $\omega_{p}^2= \omega_{\bf k}^2-\frac{k_x^2 \pi^2}{L_0^2}$.
It is interesting to compare this rate
with that for Dirichlet conditions, which is given by
$\lambda_D={1\over 2\omega_{\bf k}}(\omega_{\bf k}^2-\omega_{p}^2)$
\cite{crocce}.
We have
\begin{equation}
{\lambda_N\over\lambda_D}={\omega_{\bf k}^2+\omega_{p}^2
\over \omega_{\bf k}^2-\omega_{p}^2} > 1.
\end{equation}
For a given mode, the rate for Neumann boundary conditions is
always bigger than the rate for Dirichlet conditions.

Let us now assume the existence of one mode, say ${\bf{ j}}$,
that satisfies $\omega_{\bf j} = 3 \omega_{\bf k}$
 and $j_y=k_y,\,\, j_z=k_z$. We obtain for
$A^{({\bf n})}_{\bf k}$ and
$B^{({\bf n})}_{\bf k}$
\begin{eqnarray}
{d A^{({\bf n})}_{\bf k}\over d\tau} &=&
-\frac{1}{2 \omega_{\bf k}} \left[
\frac{k_x^2 \pi^2}{L_0^2} - 2 \omega_{\bf k}^2 \right]
B^{({\bf n})}_{\bf k}
+\frac{1}{2 \omega_{\bf k}} \left[(-1)^{k_x+j_x}
\frac{j_x^2 \pi^2}{L_0^2} - 6 \omega_{\bf k}^2 \right]
A^{({\bf n})}_{\bf j} ,
\nonumber\\
{d B^{({\bf n})}_{\bf k}\over d\tau} &=&
-\frac{1}{2 \omega_{\bf k}} \left[
\frac{k_x^2 \pi^2}{L_0^2} - 2 \omega_{\bf k}^2 \right]
A^{({\bf n})}_{\bf k}
+\frac{1}{2 \omega_{\bf k}} \left[(-1)^{k_x+j_x}
\frac{j_x^2 \pi^2}{L_0^2} - 6 \omega_{\bf k}^2 \right]
B^{({\bf n})}_{\bf j} .
\label{coupled1}
\end{eqnarray}
We also assume that the spectrum is such that the mode ${\bf {j}}$ is only
coupled to
the mode ${\bf {k}}$. The equations for $A^{({\bf n})}_{\bf j}$ and
$B^{({\bf n})}_{\bf j}$ are therefore
\begin{eqnarray}
{d A^{({\bf n})}_{\bf j}\over d\tau} &=&
-\frac{1}{6 \omega_{\bf k}} \left[(-1)^{k_x+j_x}
\frac{k_x^2 \pi^2}{L_0^2} + 2 \omega_{\bf k}^2 \right]
A^{({\bf n})}_{\bf k} ,
\nonumber\\
{d B^{({\bf n})}_{\bf j}\over d\tau} &=&
-\frac{1}{6 \omega_{\bf k}} \left[(-1)^{k_x+j_x}
\frac{k_x^2 \pi^2}{L_0^2} + 2\omega_{\bf k}^2 \right]
B^{({\bf n})}_{\bf k} .
\label{coupled2}
\end{eqnarray}
We write the system of equations in matrix form

\begin{equation}
\frac{d {\bf v}(\tau)}{d \tau}={{\cal M}}\,
{\bf v}(\tau) ,
\end{equation}
where

\begin{eqnarray}
{\bf v}(\tau)=\left(\begin{array}{c}
A^{({\bf n})}_{{\bf k}}(\tau)\\
B^{({\bf n})}_{{\bf k}}(\tau)\\
A^{({\bf n})}_{{\bf j}}(\tau)\\
B^{({\bf n})}_{{\bf j}}(\tau)
\end{array}
\right), \
{\cal M}=\frac{1}{2 \omega_{\bf k}}\left(\begin{array}{cccc}
0 & a  &
b& 0 \\
a  &
0 &0 & b \\
c&0&0&0\\
0 & c&0&0\\
\end{array}
\right) ,
\end{eqnarray}
where $a=\left[-\frac{k_x^2 \pi^2}{L_0^2} + 2 \omega_{\bf k}^2 \right]$, $b=
\left[(-1)^{k_x+j_x}
 \frac{j_x^2 \pi^2}{L_0^2} - 6 \omega_{\bf k}^2 \right]$,
 and $c=-\frac{1}{3} \left[(-1)^{k_x+j_x}
\frac{k_x^2 \pi^2}{L_0^2} + 2\omega_{\bf k}^2 \right]$ .
The solution to this system can be easily obtained after
diagonalizing the matrix
${\cal M}$. The eigenvalues are given by
\begin{equation}
\lambda = \frac{1}{4 \omega_{\bf k}}(\pm a\pm\sqrt{a^2+4bc}) .
\label{4eigen}
\end{equation}
We note that the exponential growth rate in the uncoupled case is given
by $\lambda_N=a/2\omega_{\bf k}$. When two modes are coupled, the rate
is given by the real part of the biggest eigenvalue in Eq.(\ref{4eigen}).
When $a^2+4 b c < 0$, the rate is half the one
expected for the resonant mode when the coupling is neglected, as for
Dirichlet boundary conditions \cite{crocce}. It is easy to show that
this is the case if  $(-1)^{k_x+j_x}=+1$. However, in the opposite
case, $b c>0$, and the growth rate for coupled modes is bigger
than $\lambda_N=a/2\omega_{\bf k}$ .

A relevant case where two modes are coupled is the cubic cavity
$L_x=L_y=L_z=L$.
We fix $\Omega$ as twice the lowest cavity frequency,

\begin{equation}
\Omega=2\omega_{(1,1,1)} = \frac{2 \pi \sqrt{3}}{L} .
\label{omega1}
\end{equation}
The fundamental mode ${\mathbf{k}}=(1,1,1)$ is coupled to
${\mathbf{j}}=(5,1,1)$ because $\omega_{(5,1,1)}=3\omega_{(1,1,1)}$.
Only these two modes are coupled, since there does not exist in the spectrum
any mode ${\mathbf{s}}$ satisfying $\omega_{\mathbf{s}}=5\omega_{(1,1,1)}$.
For this particular case, the four eigenvalues are

\begin{equation}
\lambda = \frac{\pi}{4 \sqrt 3 L}(\pm 5\pm 6.35 i) .
\label{4eigen2}
\end{equation}
Had we neglected the intermode coupling, we would have concluded that the
growth rate in the fundamental mode would be $\lambda= 2.5 \pi /\sqrt 3 L$.
The growth rate in the coupled case is one half of this.

One striking new feature is the possibility to enhance the exponential
growth rate by means of mode coupling, provided that the two coupled modes
satisfy $(-1)^{k_x+j_y}=-1$. As an example let us consider a cavity with
dimensions $L_y=L_z=4 L_x$. We set the external frequency to be

\begin{equation}
\Omega=2\omega_{(0,1,1)} = \frac{2 \pi}{\sqrt{8} L_x} .
\end{equation}
If this is the case, then the mode ${\bf k}=(0,1,1)$ is coupled to
${\bf j}=(1,1,1)$. The four eigenvalues are

\begin{equation}
\lambda = \frac{\sqrt{8} \pi}{16 L_x}(\pm 1 \pm \sqrt{31/3}) .
\end{equation}
This means that the exponential growth is at the rate $0.74 \pi/ L_x$,
which is more than twice the value we would had predicted  had we
neglected the coupling ($\pi/\sqrt{8} L_x$).

%%%%%%%%%%%%%%%%%%%%%%%%%%%%%%%%%%%%%%%%%%%%%%%%%%%%%%%%%%%%%%%%%%%%%%%%%

\section{The electromagnetic field}

\subsection{Transverse electric modes}

For the TE case, the expansion of the vector potential for an
arbitrary moment of time,
in terms of creation and annihilation operators, can be written as

\begin{equation}
{\bf A}^{({\rm TE})}({\bf x},t)=\sum_{{\bf n}}a^
{{\scriptscriptstyle {\rm in}}}_{{\bf n}} {\bf u}^{({\rm TE})}_{{\bf n}}
({\bf x},t) +   {\rm H.c.} .
\label{afield}
\end{equation}
For $t\leq0$ the cavity is static, and each field mode is given by

\begin{eqnarray}
{\bf u}^{({\rm TE})}_{{\bf n}} ({\bf x},t\leq0) &=&
{1\over\sqrt{2\omega_{{\bf n}}}} \sqrt{{8\over L_{x}L_{y}L_{z}}} \times
\nonumber \\
&& \left(
0,
\alpha_{{\bf n}}
\sin\left({\pi n_{x}\over L_{x}}x\right)
\cos\left({\pi n_{y}\over L_{y}}y\right)
\sin\left({\pi n_{z}\over L_{z}}z\right),
\beta_{{\bf n}} \sin\left({\pi n_{x}\over L_{x}}x\right)
\sin\left({\pi n_{y}\over L_{y}}y\right)
\cos\left({\pi n_{z}\over L_{z}}z\right)
\right) e^{-i \omega_{{\bf n}} t} ,
\label{solucionenregion`in'}
\end{eqnarray}
where $n_x$, $n_y$, and $n_z$ are integers such that $n_x \geq 1$,
$n_y,n_z \geq 0$, and $n_y, n_z$ cannot be simultaneously zero.
The constants $\alpha_{{\bf n}}$ and $\beta_{{\bf n}}$ are
components of the polarization vector for the electromagnetic field,
satisfying the normalization condition $\alpha_{{\bf n}}^2 +
\beta_{{\bf n}}^2=1$, and the Coulomb gauge condition, $\alpha_{{\bf n}}n_y/L_y
+ \beta_{{\bf n}}n_z/L_z =0$.

When $t>0$, we expand the mode functions in Eq.(\ref{afield}) with respect
to an \textit{instantaneous basis}

\begin{equation}
{\bf u}^{({\rm TE})}_{{\bf n}}({\bf x}, t>0)=
\sum_{{\bf k}} Q^{({\bf n})}_{{\bf k},{\rm TE}}(t)
\sqrt{{2\over L_x(t)}} \sin\left({\pi n_{x}\over L_{x}(t)}x\right)
{\bf \Phi}^{({\rm TE})}_{k_yk_z}(y,z),
\label{hip1}
\end{equation}
where ${\bf \Phi}^{({\rm TE})}_{k_yk_z}$ is

\begin{equation}
{\bf \Phi}^{({\rm TE})}_{k_yk_z} (y,z)=
\sqrt{{4\over L_{y}L_{z}}}
\left(
0,
\alpha_{{\bf k}}
\cos\left({\pi k_{y}\over L_{y}}y\right)
\sin\left({\pi k_{z} \over L_{z}}z\right),
\beta_{{\bf k}}
\sin\left({\pi k_{y}\over L_{y}}y\right)
\cos\left({\pi k_{z}\over L_{z}}z \right)
\right) .
\label{hip2}
\end{equation}
The functions ${\bf \Phi}^{({\rm TE})}_{k_yk_z}$ form a complete set
satisfying

\begin{equation}
\int_{0}^{L_{y}}dy\int_{0}^{L_{z}}dz \;
{\bf \Phi}^{({\rm TE})}_{k_yk_z}
\cdot{\bf \Phi}^{({\rm TE}) \star}_{j_yj_z}
=\delta_{k_yj_y}\delta_{k_zj_z}.
\end{equation}

From the above Eq.(\ref{hip1}) it is easy to obtain the dynamical
equations for the modes $Q^{({\bf n})}_{{\bf k},{\rm TE}}$. We get

\begin{equation}
\ddot{Q}^{({\bf n})}_{{\bf k},{\rm TE}}+\omega_{\bf{n}}^{2}(t)\,
Q^{({\bf n})}_{{\bf k},{\rm TE}}
=
2\lambda(t)\sum_{\bf{j}} g_{\bf{kj}}
\dot{Q}^{({\bf n})}_{{\bf j},{\rm TE}}
+\dot{\lambda}(t) \sum_{\bf{j}}g_{\bf{kj}}
 Q^{({\bf n})}_{{\bf j},{\rm TE}}(t)
\label{ecqte}
\end{equation}
where
\begin{equation}
g_{\bf{kj}}= -g_{\bf{jk}}=
\left\{
\begin{array}{ll}
(-1)^{k_{x}+j_{x}} \frac{2 k_x j_{x} }{j_{x}^{2}-k_{x}^{2}}\,
\delta_{k_{y}j_{y}}\,\delta_{k_{z}j_{z}} \,& \mbox{if $k_{x}\neq j_{x}$} ,\\
0
 & \mbox{if $k_{x}=j_{x}$}  .
\end{array}
\right.
\label{gdir}
\end{equation}
As expected, these equations are exactly those
corresponding to a scalar field satisfying Dirichlet
boundary conditions on the surfaces $x=0,L_x(t)$ \cite{crocce}.
Therefore, the number of created photons in the TE mode equals the number
of created Dirichlet scalar particles.

As an example, let us consider the parametric resonant case
$\Omega=2 \omega_{\bf k}$ for a cubic cavity. For uncoupled ${\bf k}$ modes
(such as either of the two fundamental TE modes, ${\bf k}=(1,1,0)$ and
${\bf k}=(1,0,1)$), the number of TE photons grows exponentially as
\begin{equation}
\langle {\mathcal{N}}_{{\bf k},{\rm TE}} \rangle=
\sinh^2(\lambda_D \epsilon t) ,
\label{growthD}
\end{equation}
where $\lambda_{\rm D}$  is the growth rate
for Dirichlet scalar particles, introduced in Section III.D.
For the above mentioned fundamental modes, $\lambda_D =\pi / 2 \sqrt 2 L$.
The first
coupled TE mode is ${\bf k}=(1,1,1)$, which only couples to the TE mode
${\bf j} = (5,1,1)$. At large times $\epsilon t/L \gg 1$ the number of TE photons
in those modes grows as $\langle {\cal N}_{{\bf k},{\rm TE}}\rangle \approx \langle{\cal N}_{{\bf j},{\rm TE}}\rangle\approx e^{0.9 \epsilon t/L}$ \cite{crocce}.

\subsection{Transverse magnetic modes}

The expansion in terms of creation and annihilation operators is again
of the form Eq.(\ref{afield}), but now for $t\leq0$
each field mode is given by

\begin{eqnarray}
{\bf u}^{({\rm TM})}_{{\bf n}} ({\bf x},t\leq0) &=&
{1\over\sqrt{2\omega_{{\bf n}}}} \sqrt{{8\over L_{x}L_{y}L_{z}}} \times
\nonumber \\
&& \left(
0,
\alpha_{{\bf n}}
\cos\left({\pi n_{x}\over L_{x}}x\right)
\sin\left({\pi n_{y}\over L_{y}}y\right)
\cos\left({\pi n_{z}\over L_{z}}z\right),
\beta_{{\bf n}} \cos\left({\pi n_{x}\over L_{x}}x\right)
\cos\left({\pi n_{y}\over L_{y}}y\right)
\sin\left({\pi n_{z}\over L_{z}}z\right)
\right) e^{-i \omega_{{\bf n}} t} .
\label{solucionenregion`in'tm}
\end{eqnarray}
Here $n_x$, $n_y$, and $n_z$ are non-negative integers, and $n_y$ and $n_z$
cannot be simultaneously zero.

On the other hand, when $t>0$ we introduce an instantaneous basis similar
to that of the scalar field in Section III. We write

\begin{equation}
{\bf u}^{({\rm TM})}_{{\bf n}}({\bf x}, t>0)=
\sum_{{\bf k}}\left (Q^{({\bf n})}_{{\bf k},{\rm TM}}(t)
+ \dot Q^{({\bf n})}_{{\bf k},{\rm TM}}(t) g(x,t)\right) \sqrt{{2\over L_x(t)}}
\cos\left({\pi n_{x}\over L_{x}(t)}x\right)
{\bf \Phi}^{({\rm TM})}_{k_yk_z}(y,z),
\label{hiperimportante1tm}
\end{equation}
where the functions  ${\bf \Phi}^{({\rm TM})}_{k_yk_z}$
are similar to their TE counterparts (they can be obtained by interchanging
$\cos$ and $\sin$ in the RHS of Eq.(\ref{hip2}).
Since all TE modes have $n_x=0$, the first mode of the cavity that can be
excited by the external frequency is a TM mode. In particular, 
for a cavity such that $L_x \ll L_y,L_z$ only TM modes can be excited.

From the above equation, it is now clear that the dynamical evolution of the
TM modes is that of a scalar field satisfying generalized Neumann boundary
conditions. As a consequence,
the number of created photons in the TM mode equals the number of created
Neumann scalar particles. Again, let us consider the parametric resonant case
$\Omega=2 \omega_{\bf k}$ for a cubic cavity. For uncoupled ${\bf k}$ modes
(such as either of the two fundamental TM modes, ${\bf k}=(0,1,0)$ and
${\bf k}=(0,0,1)$), the number of TM photons grows exponentially as

\begin{equation}
\langle {\mathcal{N}}_{{\bf k},{\rm TM}} \rangle=
\sinh^2(\lambda_N \epsilon t) ,
\label{growthN}
\end{equation}
where $\lambda_N$ is the growth rate for Neumann scalar particles, also
introduced in Section III.D. For the fundamental modes 
$\lambda_N=\pi/L$. The first coupled TM mode is ${\bf k}=(0,1,1)$,
which only couples to the TM mode ${\bf j}=(4,1,1)$. For large times ($\epsilon t/L \gg 1$) 
the number of particles in these modes grows as 
$\langle {\mathcal{N}}_{{\bf k},{\rm TM}} \rangle \approx
\langle {\mathcal{N}}_{{\bf j},{\rm TM}} \rangle \approx e^{4.4 \epsilon t/L}$.
The next coupled TM mode is the
same as the TE mode, namely ${\bf k}=(1,1,1)$, coupled to ${\bf j}=(5,1,1)$.
The exponential growth is 
$\langle {\mathcal{N}}_{{\bf k},{\rm TM}} \rangle \approx
\langle {\mathcal{N}}_{{\bf j},{\rm TM}} \rangle \approx e^{4.5 \epsilon t/L}$,
the growth rate for these modes being greater than that for the TE case.

%%%%%%%%%%%%%%%%%%%%%%%%%%%%%%%%%%%%%%%%%%%%%%%%%%%%%%%%%%%%%%%%%%%%%%%%

\section{DISCUSSION}

In this paper we have computed the resonant photon creation inside a
three dimensional oscillating cavity taking the
vector nature of the electromagnetic field into account.
Previous works studied the case of a scalar field with Dirichlet
boundary conditions. As the electromagnetic field involves both
Neumann and Dirichlet boundary conditions, we first analyzed a
massless scalar field satisfying (generalized) Neumann boundary
conditions. We have shown that in this case it is also possible
to expand the field modes in terms of an instantaneous basis, the difference
with the Dirichlet case being that the expansion is not unique - it depends
on an arbitrary function $g(x,t)$ satisfying the boundary conditions
Eq.(\ref{bc}). However, physical quantities like the number of created
particles or the energy density inside the cavity are independent of the choice
of such a function.
After treating the Neumann scalar case we considered the full electromagnetic
problem and showed that the TE modes of the elecromagnetic field
are essentially described by a Dirichlet scalar field, while the
TM modes correspond to a Neumann scalar field.

We have studied in detail the resonant situation
$\Omega = 2\omega_{\bf  {k}}$ for two cases: an uncoupled resonant mode and
two coupled resonant modes. In both cases, the exponential
growth  of created photons is greater for TM modes.
For the uncoupled case, we have found that
\begin{equation}
{\lambda_{{\rm TM}}\over\lambda_{{\rm TE}}}={\omega_{\bf k}^2+\omega_{p}^2
\over \omega_{\bf k}^2-\omega_{p}^2}.
\end{equation}
For a cavity with $L_x\simeq L_y\simeq L_z$, $\omega_{p}^2\simeq
\frac{2}{3}\omega_{\bf k}^2$ so $\lambda_{{\rm TM}}\simeq 5\lambda_{{\rm TE}}$.
We can estimate the number of created TE and TM photons given
by Eqs.(\ref{growthD},\ref{growthN}) using typical values for the maximal dimensionless 
displacement $\epsilon$ that may be obtained in conceivable future experiments.
For 3D cubic cavities of linear dimensions of the order of $1-10$ cm, the lowest
resonant frequency is of the order of GHz. It may turn out to be very difficult, 
if not impossible, to make the cavity oscillate as a whole at such a high frequency.
To overcome this difficulty a different experimental scenario was proposed in \cite{dodexp}, 
consisting of strong acoustic waves excited on the surface of the cavity wall. 
Typical materials cannot bear relative amplitude deformations in excess of 
$\delta_{\rm max}=10^{-2}$. This sets a limit to the maximum velocity of the boundary,
$v_{\rm max}=\delta_{\rm max} v_s \approx 50 {\rm m/s},$ 
($v_s$ is the speed of sound in the material), and consequently to the maximal
dimensionless displacement $\epsilon_{\rm max} = v_{\rm max}/\Omega L$. For example,
for a cavity with $L=10$cm whose lowest mode (i.e., either of the two TM modes
${\bf k}=(0,1,0)$ or ${\bf k}=(0,0,1)$) is being excited ($\Omega=2\pi c/L=18$GHz),
we get $\epsilon_{\rm max} \approx  10^{-8}$. Even for a value of $\epsilon$ 10 times
smaller than this, one gets an exponentially large number of created photons 
$\langle {\cal N}_{{\bf k}, {\rm TM}} \rangle = \sinh^2(10 t/{\rm s})$ which, after 1 second,
gives a total of approximately $10^8$ photons created in that mode. We can also compare
the number of photons produced for an uncoupled mode ${\bf k}$, common to both kind of polarizations
TE and TM. For example, for the mode ${\bf k}=(1,1,0)$ one gets 
$\langle {\cal N}_{{\bf k},{\rm TE}} \rangle \approx \sinh^2(3 t /{\rm s})$ and
$\langle {\cal N}_{{\bf k},{\rm TM}} \rangle \approx \sinh^2(10 t /{\rm s})$, which
after 1 second produces a total of $10^2$ TE photons and $10^8$ TM photons.
For the case of two coupled modes we have found that, for Neumann boundary
conditions, the coupling can enhance the exponential growth. This contrasts
with the case of Dirichlet boundary conditions, in which the coupling always
suppresses the exponential growth. These facts may be relevant for an
eventual experimental verification of the dynamical Casimir effect.

All the above considerations assume ideal conditions, such as perfectly conducting
plates, exact parametric resonant condition $\Omega=2 \omega_{\bf k}$,
arbitrary large $Q$ factor for the cavity (no leakage of photons), no thermal
noise, etc. Some of these conditions were relaxed in our previous paper
\cite{crocce}, where we analyzed, for Dirichlet boundary conditions,
the enhancement of photon creation due to finite temperature effects,
slightly off-resonance situations, the case of three coupled modes, etc.
The generalization of these findings to the electromagnetic case is not
too complicated, and we expect similar conclusions. Given our results
for the dynamical behavior of TE and TM modes,
it is also possible to study the full electromagnetic problem in three
dimensional leaky cavities along the lines of \cite{leaky}.

%%%%%%%%%%%%%%%%%%%%%%%%%%%%%%%%%%%%%%%%%%%%%%%%%%%%%%%%%%%%%%%%%%%%%%%%%%%%

\section{ACKNOWLEDGEMENTS}

We are grateful to R. Onofrio for useful comments. The  work of FDM  was
supported by Universidad de Buenos Aires, Conicet,
and Agencia Nacional de Promoci\'on Cient\'\i fica y Tecnol\'ogica, Argentina.

%%%%%%%%%%%%%%%%%%%%%%%%%%%%%%%%%%%%%%%%%%%%%%%%%%%%%%%%%%%%%%%%%%%%%%%%%

\end{document}